\documentclass[a4]{article}
\usepackage{spconf,amsmath,graphicx,amsfonts}
\usepackage{regensky}

\usepackage[firstpage=true]{background}

\SetBgContents{\parbox{\textwidth}{\footnotesize © 2023 IEEE. Personal use of this material is permitted. Permission from IEEE must be
obtained for all other uses, in any current or future media, including
reprinting/republishing this material for advertising or promotional purposes, creating new
collective works, for resale or redistribution to servers or lists, or reuse of any copyrighted
component of this work in other works.}}
\SetBgScale{1}
\SetBgAngle{0}
\SetBgPosition{current page.north}
\SetBgVshift{-1.6cm}
\SetBgColor{black}
\SetBgOpacity{1}

\title{Motion Plane Adaptive Motion Modeling\\for Spherical Video Coding in H.266/VVC}

\name{Andy Regensky, Christian Herglotz, André Kaup\thanks{The authors gratefully acknowledge that this work has been supported by the Deutsche Forschungsgemeinschaft (DFG, German Research Foundation) under project number 418866191.}}
\address{Friedrich-Alexander-Universität Erlangen-Nürnberg\\Multimedia Communications and Signal Processing\\Cauerstr. 7, 91058 Erlangen, Germany}

\begin{document}
\maketitle

\begin{abstract}
  Motion compensation is one of the key technologies enabling the high compression efficiency of modern video coding standards.
  To allow compression of spherical video content, special mapping functions are required to project the video to the 2D image plane.
  Distortions inevitably occurring in these mappings impair the performance of classical motion models.
  In this paper, we propose a novel motion plane adaptive motion modeling technique (MPA) for spherical video that allows to perform motion compensation on different motion planes in 3D space instead of having to work on the - in theory arbitrarily mapped - 2D image representation directly.
  The integration of MPA into the state-of-the-art H.266/VVC video coding standard shows average Bjøntegaard Delta rate savings of 1.72\% with a peak of 3.37\% based on PSNR and 1.55\% with a peak of 2.92\% based on WS-PSNR compared to VTM-14.2.
\end{abstract}


\section{Introduction}\label{sec:introduction}

With the increasing availability of affordable VR headsets, the demand for spherical 360-degree video has the potential to rise significantly within the coming years.
Especially due to the high required frame rates and resolutions, the storage and distribution of immersive 360-degree VR experiences is expensive.
The development and design of improved compression techniques for spherical video is thus indispensable.

Typically, a 2D representation of the 360-degree video is required to allow compression using existing video coding techniques such as
the H.265/HEVC~\cite{Sullivan2012} or the H.266/VVC~\cite{Bross2021a} video coding standard.
\begin{figure}
  \centering
  \subfloat[Projected spherical video frame\label{fig:360image:erp}]{%
  \includegraphics[width=0.64\linewidth]{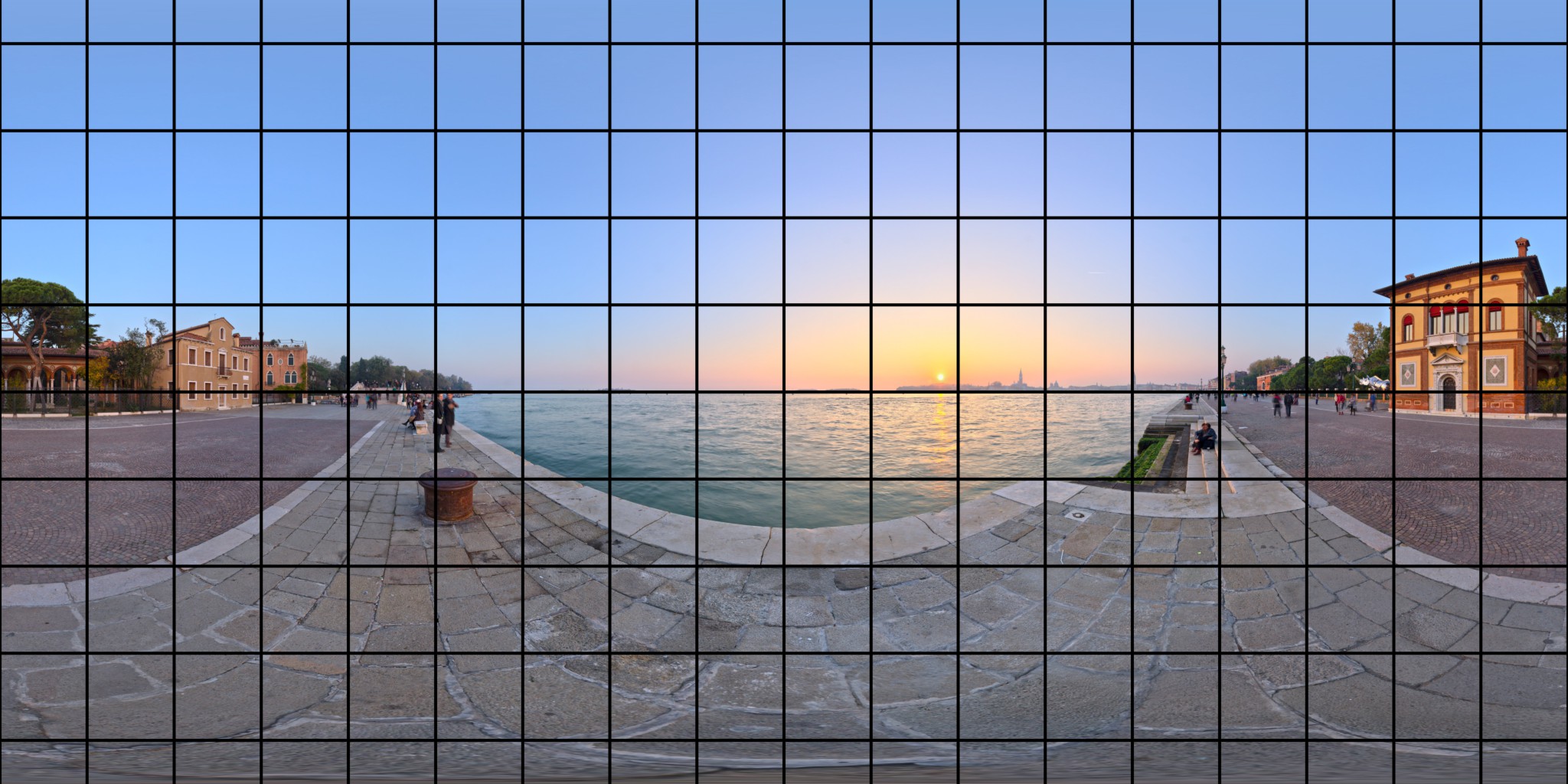}}
  \hfill%
  \subfloat[Virtual camera\label{fig:360image:front}]{%
  \input{tikz/sphere_camera_front.tex}}

  \caption{(a) Spherical video mapped to the 2D image plane using an equirectangular projection. (b) Visualization of the concept of motion planes, where a virtual perspective camera is positioned in the 360-degree video sphere.}
  \label{fig:360image}
\end{figure}
Compared to coding conventional perspective video, the compression performance of modern video codecs is notably reduced for non-perspective projection formats~\cite{Bross2021a, Wien2017}.
One reason for this are the distortions that inevitably occur due to the mapping of the spherical 360-degree video to the 2D image plane~\cite{Pearson1990} as visible in Fig.~\ref{fig:360image:erp}.

In this work, we propose the integration of a novel \textit{motion plane adaptive} motion model (MPA) for spherical video into the H.266/VVC video coding standard.
The model allows to perform motion compensation on different motion planes in 3D space, while any motion on these planes is modeled purely using horizontal and vertical shifts.
Fig.~\ref{fig:360image:front} offers an intuitive understanding of this concept, where a virtual perspective camera is placed in the spherical 360-degree video.
The image as seen by this virtual perspective camera represents the motion plane, whose orientation matches the orientation of the camera.
Any motion perpendicular to the camera normal can be perfectly represented using a translational motion model on the corresponding motion plane.

The remainder of the paper is organized as follows.
In Section~\ref{sec:background}, the traditional inter prediction procedure is briefly recapitulated and an overview over related motion models for spherical video coding is given.
The proposed motion modeling procedure and its integration into VVC is explained in Section~\ref{sec:proposal}.
Section~\ref{sec:performance}
evaluates the performance of MPA presenting both numerical and visual results.
Section~\ref{sec:conclusion} concludes the paper.

\section{Background and Related Work}\label{sec:background}

Motion compensation or inter prediction is a key component of any modern hybrid video codec, where a prediction is formed based on an already coded reference frame, such that only the difference between the predicted block and the actual block needs to be signaled.
Using the translational motion model, a motion compensated or predicted frame $\vec{I}_\text{pred}$ can be obtained both at the encoder and at the decoder by extracting the motion compensated pixel values from the reference frame $\vec{I}_\text{ref}$ as
\begin{align}
  \vec{I}_\text{pred}(\vec{p}) = \vec{I}_\text{ref}(\vec{p} + \vec{t}) = \vec{I}_\text{ref}(\vec{m}_\text{t}(\vec{p}, \vec{t}))~~\forall~\vec{p} \in \mathcal{B} \label{eq:tmc}
\end{align}
for all blocks in the image, where $\mathcal{B}$ denotes the set of all pixel coordinates $\vec{p} \in \mathbb{R}^2$ within the regarded block, $\vec{t} \in \mathbb{R}^2$ denotes the regarded block's motion vector, and $\vec{m}_\text{t} : \mathbb{R}^2 \times \mathbb{R}^2 \rightarrow \mathbb{R}^2$ describes the translational motion model.
Internally, a suitable interpolation method is required in order to access pixel values at fractional pixel positions.
For each block, the encoder searches the motion vector $\vec{t}$ yielding the highest compression efficiency in a process called rate-distortion optimization~\cite{Sullivan1998}.

To improve the compression efficiency of video codecs for spherical video, major efforts have gone into designing advanced motion models based on the known spherical characteristics.
In~\cite{Li2017, Li2019}, \textit{Li et al.} propose a 3D translational motion model with reported rate savings of 4.3\% for HEVC on average, where all pixels in a block are shifted in 3D space according to a 3D motion vector derived from the original 2D motion vector.
In~\cite{DeSimone2017}, \textit{De Simone et al.} propose a motion model, where the 2D motion vector is applied on a plane tangential to the block center in the spherical domain representation using a gnomonic projection.
In~\cite{Vishwanath2017, Vishwanath2018a}, \textit{Vishwanath et al.} introduce a rotational motion model with reported rate savings of 11.4\% for HEVC on average, where all pixels in a block are rotated on the sphere according to a 2D motion vector that is interpreted as rotation angles.
In~\cite{Vishwanath2018}, they furthermore propose a motion model with reported rate savings of 23.5\% for HEVC on average, where all pixels in a block are rotated along
geodesics corresponding to a known global camera motion vector.
In~\cite{Marie2021}, \textit{Marie et al.} propose a combination of the rotational and tangential motion models.

\section{Motion Plane Adaptive\newline Motion Modeling}\label{sec:proposal}

\subsection{Projections}\label{subsec:projections}

Any valid projection function~$\boldsymbol{\xi} : \mathcal{S} \rightarrow \mathbb{R}^2$ is invertible and describes the relation between a 3D space coordinate $\vec{s} = (x, y, z)^T \in \mathcal{S}$ on the unit sphere and the corresponding pixel coordinate $\vec{p} = (u, v)^T \in \mathbb{R}^2$ on the 2D image plane, where $\mathcal{S} = \{ \vec{s} \in \mathbb{R}^3\ |\ \lVert\vec{s}\rVert_2 = 1 \}$ describes the set of all coordinates on the unit sphere.

The equirectangular projection $\boldsymbol{\xi}_\text{erp}: \mathcal{S} \rightarrow \mathbb{R}^2$ is a popular and widely applied example of a general 360-degree projection.
It maps the polar angle $\theta \in [0, \pi]$ to the vertical $v$-axis and the azimuthal angle $\varphi \in [0, 2\pi]$ to the horizontal $u$-axis of the 2D image plane.
The inverse equirectangular projection $\boldsymbol{\xi}_\text{erp}^{-1}: \mathbb{R}^2 \rightarrow \mathcal{S}$ reverses this procedure and projects the pixel coordinate on the 2D image plane back to the unit sphere.

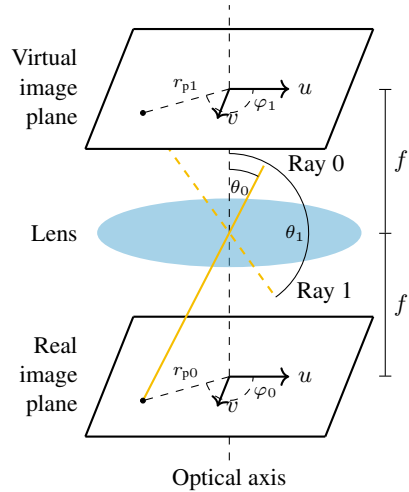
\begin{figure}[t]
  \centering
  \begin{tikzpicture}[scale=3.19]
    \small
    \definecolor{better_yellow}{HTML}{F6BE00}
    \pgfmathsetmacro{\rayalphaa}{1.4}
    \pgfmathsetmacro{\rayalphab}{1.5}

    \coordinate (O) at (0, 0);

    \path (O) +(0, -0.6) coordinate (rip_O);
    \path (rip_O) +(-0.6, -0.25) coordinate (rip_bl);
    \path (rip_bl) +(1.0, 0) coordinate (rip_br);
    \path (rip_O) +(0.6, 0.25) coordinate (rip_tr);
    \path (rip_tr) +(-1.0, 0) coordinate (rip_tl);
    \path (rip_O) +(-0.36, -0.1) coordinate (rip_intersection);
    \path (O) +($(rip_intersection) - \rayalphaa*(rip_intersection)$) coordinate (rip_ray_start);

    \path (O) +(0, 0.6) coordinate (vip_O);
    \path (vip_O) +(-0.6, -0.25) coordinate (vip_bl);
    \path (vip_bl) +(1.0, 0) coordinate (vip_br);
    \path (vip_O) +(0.6, 0.25) coordinate (vip_tr);
    \path (vip_tr) +(-1.0, 0) coordinate (vip_tl);
    \path (vip_O) +(-0.36, -0.1) coordinate (vip_intersection);
    \path (O) +($(vip_intersection) - \rayalphab*(vip_intersection)$) coordinate (vip_ray_start);

    \draw[thick] (rip_bl) -- (rip_br) coordinate[midway] (rip_edge_bottom);
    \draw[thick] (rip_br) -- (rip_tr) coordinate[midway] (rip_edge_right);
    \draw[thick] (rip_tr) -- (rip_tl);
    \draw[thick] (rip_tl) -- (rip_bl) coordinate[midway] (rip_edge_left);
    \path (rip_O) -- (rip_edge_right) coordinate[midway] (rip_x);
    \path (rip_O) -- (rip_edge_bottom) coordinate[midway] (rip_y);
    \draw[thick,->] (rip_O) -- (rip_x) node[anchor=west] {$u$};
    \draw[thick,->] (rip_O) -- (rip_y) node[anchor=west] {$v$};

    \path (rip_O |- rip_bl)+(0, -0.1) coordinate (optical_axis_bottom);
    \draw[dashed] (rip_O |- rip_bl) -- (optical_axis_bottom) node[anchor=north] {Optical axis};
    \draw[dashed] (rip_O) -- (O);
    \node[ellipse, fill=carolinablue, opacity=0.5, minimum width=100pt, minimum height=26pt] (lens) at (O) {};
    \node[anchor=east, xshift=-4pt] at (lens.west) {Lens};
    \draw[dashed] (O) -- (vip_O |- vip_bl);

    \draw[thick, better_yellow] (rip_ray_start) -- (rip_intersection);
    \draw[thick, dashed, better_yellow] (vip_ray_start) -- (vip_intersection);
    \node[anchor=west, xshift=6pt] at (rip_ray_start) {Ray 0};
    \node[anchor=west, xshift=6pt] at (vip_ray_start) {Ray 1};

    \fill[fill=white] (vip_bl)--(vip_br)--(vip_tr)--(vip_tl)--cycle;
    \draw[thick] (vip_bl) -- (vip_br) coordinate[midway] (vip_edge_bottom);
    \draw[thick] (vip_br) -- (vip_tr) coordinate[midway] (vip_edge_right);
    \draw[thick] (vip_tr) -- (vip_tl);
    \draw[thick] (vip_tl) -- (vip_bl) coordinate[midway] (vip_edge_left);
    \path (vip_O) -- (vip_edge_right) coordinate[midway] (vip_x);
    \path (vip_O) -- (vip_edge_bottom) coordinate[midway] (vip_y);
    \draw[thick,->] (vip_O) -- (vip_x) node[anchor=west] {$u$};
    \draw[thick,->] (vip_O) -- (vip_y) node[anchor=west] {$v$};

    \node[anchor=east, xshift=-4pt, align=right] at (rip_edge_left.center -| lens.west) {Real\\image\\plane};
    \node[anchor=east, xshift=-4pt, align=right] at (vip_edge_left.center -| lens.west) {Virtual\\image\\plane};

    \path (vip_O |- vip_tl)+(0, 0.1) coordinate (optical_axis_top);
    \draw[dashed] (vip_O) -- (optical_axis_top);

    \filldraw[black] (rip_intersection) circle (0.01);
    \draw[thin, dashed] (rip_O) -- (rip_intersection) node[midway, anchor=south] (rip_r) {\scriptsize$r_{\text{p}0}$};
    \draw pic[draw=black, angle radius=9pt, dashed] {angle=rip_intersection--rip_O--rip_x};
    \node[below right = 6pt and 6pt of rip_O, anchor=west] (rip_phi) {\scriptsize$\varphi_{0}$};
    \draw pic[draw=black, angle radius=24pt] {angle=rip_ray_start--O--vip_O};
    \node[above right= 11pt and 4.2pt of O, anchor=south] (rip_theta) {\scriptsize$\theta_0$};

    \filldraw[black] (vip_intersection) circle (0.01);
    \draw[thin, dashed] (vip_O) -- (vip_intersection) node[midway, anchor=south] (vip_r) {\scriptsize$r_{\text{p}1}$};
    \draw pic[draw=black, angle radius=9pt, dashed] {angle=vip_intersection--vip_O--vip_x};
    \node[below right = 6pt and 6pt of vip_O, anchor=west] (vip_phi) {\scriptsize$\varphi_{1}$};
    \draw pic[draw=black, angle radius=30pt] {angle=vip_ray_start--O--vip_O};
    \node[right = 32pt of O, anchor=east] (vip_theta) {\scriptsize$\theta_{1}$};

    \path (rip_tr) +(0.05, -0.25) coordinate (f_b);
    \path (f_b) +(0, 0.6) coordinate (f_t);
    \path (f_t) +(0, 0.6) coordinate (f_tt);
    \draw[|-|] (f_b) -- (f_t) node[midway, anchor=west] (focal_length) {$f$};
    \draw[-|] (f_t) -- (f_tt) node[midway, anchor=west] (focal_length) {$f$};
\end{tikzpicture}
  \vspace*{-2em}
  \caption{Perspective image planes. Light rays with incident angles $\theta < \pi/2$ are projected to the real image plane, while light rays with incident angles $\theta > \pi/2$ are projected to the virtual image plane.}
  \label{fig:perspective-image-planes}
\end{figure}

\begin{figure*}[t]
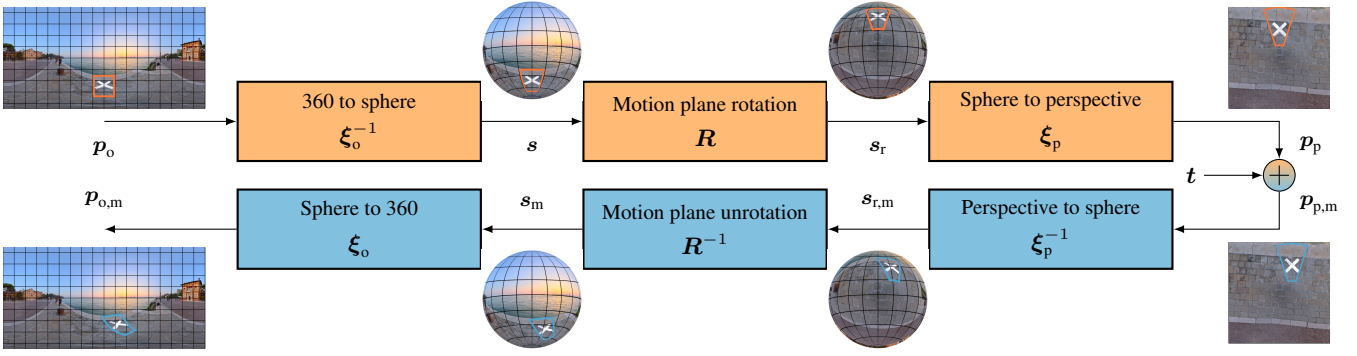

  \centering
  \include{tikz/mpa-schematic}
  \vspace*{-2em}
  \caption{Schematic representation of the motion plane adaptive motion model with a visualization of the procedure for an exemplary block in a 360-degree image employing an ERP projection. For visualization, block motion for an exemplary motion vector and rotation matrix is shown. It is clearly visible that the proposed motion model is able to accurately replicate the distortions of the block in the ERP domain resulting from a translational motion on the bottom plane.}
  \label{fig:mpa-schematic}
\end{figure*}

To map data from the unit sphere to a desired motion plane, the perspective projection is applied.
As the perspective projection is principally only defined for incident angles $\theta < \pi/2$, we propose a generalized perspective projection $\boldsymbol{\xi}_\text{p}: \mathcal{S} \rightarrow \mathbb{R}^2$ that projects light rays corresponding to incident angles $\theta > \pi/2$ to a so-called virtual image plane on the opposite side of the lens as visualized in Fig.~\ref{fig:perspective-image-planes}.
Assuming that the optical axis of the perspective projection aligns with the z-axis in 3D space, the incident angle $\theta$ and the azimuthal angle $\varphi$ of a light ray corresponding to a pixel coordinate $\vec{s}$ on the unit sphere can be directly obtained from its representation in spherical coordinates $(\rho, \theta, \varphi)$.
The intersection radius $r_\text{p}$ of the light ray with the real or virtual image plane is then obtained through
\begin{align}
  r_{\text{p}} = \begin{cases}
                   f\tan(\theta) & \text{if $0 \leq \theta < \pi/2$,} \\
                   f\tan(\pi - \theta) & \text{if $\pi/2 < \theta \leq \pi$,}
  \end{cases}\label{eq:perspective:radius}
\end{align}
and the corresponding pixel coordinate $\vec{p}_\text{p}$ is obtained by converting the polar coordinates $(r_\text{p}, \varphi$) of the intersection point to cartesian coordinates.
The inverse projection function $\boldsymbol{\xi}_\text{p}^{-1}: \mathbb{R}^2 \rightarrow \mathcal{S}$ inverts this procedure and yields the pixel coordinate $\vec{s}$ on the unit sphere based on the pixel coordinate $\vec{p}_\text{p}$ on the perspective image plane.

\subsection{Motion Model}\label{subsec:model}

In a first step, the original 360-degree pixel coordinate $\vec{p}_\text{o}$ (in, e.g., the ERP domain) is projected to the pixel coordinate $\vec{p}_\text{p}$ on the desired motion plane described by $\mat{R}$ using the reprojection function $\boldsymbol{\zeta}_{\mat{R}}: \mathbb{R}^2 \rightarrow \mathbb{R}^2$ defined as
\begin{align}
  \vec{p}_\text{p} = \boldsymbol{\zeta}_{\mat{R}}(\vec{p}_\text{o}) = \boldsymbol{\xi}_\text{p}\left( \mat{R}\boldsymbol{\xi}_\text{o}^{-1}(\vec{p}_\text{o}) \right),\label{eq:mpa-motion-model:step1}
\end{align}
where $\boldsymbol{\xi}_\text{o}^{-1}$ projects the original 360-degree pixel coordinate $\vec{p}_\text{o}$ to the unit sphere, the motion plane rotation matrix $\mat{R} \in \mathbb{R}^{3 \times 3}$ rotates the pixel coordinate on the unit sphere according to the desired motion plane orientation, and $\boldsymbol{\xi}_\text{p}$ then projects the pixel coordinate onto the motion plane.

In a second step, the translational motion according to the motion vector $\vec{t}$ is performed on the obtained motion plane yielding the moved pixel coordinate
\begin{align}
  \vec{p}_\text{p,m} = \vec{p}_\text{p} + \vec{t}.\label{eq:mpa-motion-model:step2}
\end{align}

In the final third step, the moved pixel coordinate $\vec{p}_\text{p,m}$ on the motion plane is projected back to the original 360-degree format to obtain the moved pixel coordinate in the 360-degree projection $\vec{p}_\text{o,m}$ using the inverse reprojection function $\boldsymbol{\zeta}_{\mat{R}}^{-1}$ yielding
\begin{align}
  \vec{p}_\text{o,m} = \boldsymbol{\zeta}^{-1}_{\mat{R}}(\vec{p}_\text{p,m}) = \boldsymbol{\xi}_\text{o}\left( \mat{R}^{-1}\boldsymbol{\xi}_\text{p}^{-1}(\vec{p}_\text{p,m}) \right).\label{eq:mpa-motion-model:step3}
\end{align}

Combining steps~\eqref{eq:mpa-motion-model:step1}-\eqref{eq:mpa-motion-model:step3}, the overall motion model $\vec{m}_\text{mpa}$ is defined as
\begin{align}
  \vec{m}_\text{mpa}(\vec{p}_\text{o}, \vec{t}, \mat{R}) &= \boldsymbol{\zeta}^{-1}_{\mat{R}}\big(\boldsymbol{\zeta}_{\mat{R}}(\vec{p}_\text{o}) + \vec{t}\big) \label{eq:mpa-motion-model} 
\end{align}
A schematic representation of the described motion model is shown in Fig.~\ref{fig:mpa-schematic}.
For application of MPA in video coding, we formulate a limited set of three rotation matrices yielding the motion planes \textit{front/back}, \textit{left/right}, and \textit{top/bottom} that are selectable on a per CU basis.

\begin{table*}
  \centering
  \renewcommand{\arraystretch}{1.1}
  \caption{Comparison of BD-Rate in \% with respect to VTM-14.2 for different 360-degree motion models based on WS-PSNR. Negative values (black) represent actual rate savings, positive values (red) represent increases in rate. Bold entries mark the hightest rate savings.}\label{table:models-table}
  \small
  \begin{tabular}{l||r|r|r|r|r|r||r|r|r|r|r|r}
& \multicolumn{6}{c||}{PSNR} & \multicolumn{6}{c}{WS-PSNR} \\ 
\hline 
\multicolumn{1}{r||}{} & \multicolumn{1}{c|}{\makecell[c]{MPA}} & \multicolumn{1}{c|}{\makecell[c]{3DT\\\cite{Li2017,Li2019}}} & \multicolumn{1}{c|}{\makecell[c]{TAN\\\cite{DeSimone2017}}} & \multicolumn{1}{c|}{\makecell[c]{ROT\\\cite{Vishwanath2017,Vishwanath2018a}}} & \multicolumn{1}{c|}{\makecell[c]{GED\\\cite{Vishwanath2018}}} & \multicolumn{1}{c||}{\makecell[c]{TARO\\\cite{Marie2021}}} & \multicolumn{1}{c|}{\makecell[c]{MPA}} & \multicolumn{1}{c|}{\makecell[c]{3DT\\\cite{Li2017,Li2019}}} & \multicolumn{1}{c|}{\makecell[c]{TAN\\\cite{DeSimone2017}}} & \multicolumn{1}{c|}{\makecell[c]{ROT\\\cite{Vishwanath2017,Vishwanath2018a}}} & \multicolumn{1}{c|}{\makecell[c]{GED\\\cite{Vishwanath2018}}} & \multicolumn{1}{c}{\makecell[c]{TARO\\\cite{Marie2021}}} \\
\hline 
SkateboardInLot & \textbf{\textcolor{black}{-0.97}}  & \textcolor{red}{0.32}  & \textcolor{red}{0.26}  & \textcolor{red}{0.33}  & \textcolor{red}{0.47}  & \textcolor{red}{0.05}  & \textbf{\textcolor{black}{-1.06}}  & \textcolor{red}{0.13}  & \textcolor{red}{0.03}  & \textcolor{black}{-0.01}  & \textcolor{black}{-0.13}  & \textcolor{black}{-0.11} \\ 
ChairliftRide & \textcolor{black}{-1.56}  & \textcolor{black}{-0.73}  & \textcolor{black}{-1.02}  & \textcolor{black}{-0.68}  & \textbf{\textcolor{black}{-1.65}}  & \textcolor{black}{-0.95}  & \textbf{\textcolor{black}{-1.22}}  & \textcolor{black}{-0.16}  & \textcolor{black}{-0.53}  & \textcolor{black}{-0.15}  & \textcolor{black}{-1.19}  & \textcolor{black}{-0.58} \\ 
Balboa & \textbf{\textcolor{black}{-2.05}}  & \textcolor{black}{-0.13}  & \textcolor{black}{-0.23}  & \textcolor{black}{-0.05}  & \textcolor{black}{-0.64}  & \textcolor{black}{-0.25}  & \textbf{\textcolor{black}{-1.87}}  & \textcolor{black}{-0.07}  & \textcolor{black}{-0.12}  & \textcolor{red}{0.03}  & \textcolor{black}{-0.53}  & \textcolor{black}{-0.15} \\ 
Broadway & \textbf{\textcolor{black}{-1.63}}  & \textcolor{red}{0.10}  & \textcolor{black}{-0.04}  & \textcolor{red}{0.14}  & \textcolor{black}{-0.63}  & \textcolor{black}{-0.08}  & \textbf{\textcolor{black}{-1.64}}  & \textcolor{red}{0.11}  & \textcolor{black}{-0.00}  & \textcolor{red}{0.13}  & \textcolor{black}{-0.61}  & \textcolor{black}{-0.08} \\ 
Landing2 & \textcolor{black}{-3.37}  & \textcolor{black}{-1.23}  & \textcolor{black}{-2.86}  & \textcolor{black}{-0.95}  & \textbf{\textcolor{black}{-4.03}}  & \textcolor{black}{-1.99}  & \textcolor{black}{-2.92}  & \textcolor{black}{-1.03}  & \textcolor{black}{-2.54}  & \textcolor{black}{-0.83}  & \textbf{\textcolor{black}{-3.50}}  & \textcolor{black}{-1.82} \\ 
BranCastle2 & \textcolor{black}{-0.74}  & \textcolor{black}{-0.32}  & \textcolor{black}{-0.47}  & \textcolor{black}{-0.30}  & \textbf{\textcolor{black}{-0.77}}  & \textcolor{black}{-0.36}  & \textcolor{black}{-0.58}  & \textcolor{black}{-0.22}  & \textcolor{black}{-0.43}  & \textcolor{black}{-0.21}  & \textbf{\textcolor{black}{-0.62}}  & \textcolor{black}{-0.30} \\ 
\hline 
Average & \textbf{\textcolor{black}{-1.72}}  & \textcolor{black}{-0.33}  & \textcolor{black}{-0.72}  & \textcolor{black}{-0.25}  & \textcolor{black}{-1.21}  & \textcolor{black}{-0.60}  & \textbf{\textcolor{black}{-1.55}}  & \textcolor{black}{-0.21}  & \textcolor{black}{-0.60}  & \textcolor{black}{-0.17}  & \textcolor{black}{-1.09}  & \textcolor{black}{-0.51} \\ 
\end{tabular}

\end{table*}

\subsection{Codec Integration}\label{subsec:codec}

The proposed MPA is integrated into the state-of-the-art H.266/VVC video coding standard~\cite{Bross2021a} as an additional tool\footnote{The source code of our MPA-VVC implementation is publicly available at \textit{https://github.com/fau-lms/vvc-extension-mpa}}.
The implementation is based on the VVC reference software VTM-14.2~\cite{Browne2021, VTM-14.2}.
For each CU, an additional flag signals whether MPA is enabled.
If MPA is enabled, the employed motion plane is signaled through a flag denoting whether the front/back motion plane is used, and an optional second flag denoting whether the left/right or the top/bottom motion plane is used.
All flags are coded using CABAC~\cite{Marpe2003} with dedicated context models.
In case of bi-prediction, both prediction directions share the same motion plane.
To reduce the computational complexity of MPA, the motion modeling procedure is executed on $4 \times 4$ subblocks, where the resulting pixel shift is calculated for each subblock center and then shared among all pixels within the same subblock.
Any required signal interpolations use the existing interpolation filters available in VVC~\cite{VVC-Draft10}.
The encoder selects the best suitable motion plane by performing a conventional motion search on each motion plane and selecting the motion plane yielding the lowest rate-distortion cost.
If the obtained rate-distortion cost of MPA is lower than that of the classical translational and affine inter prediction pipeline, MPA is selected for the given CU.
Compatibility with all merge modes, decoder-side motion vector refinement, bi-directional optical flow, adaptive motion vector resolution, bi-prediction with CU-level weights and symmetric motion vector difference signaling is ensured.

\section{Performance Evaluation}\label{sec:performance}

We evaluate our approach on the luminance channel of six ERP sequences with strong motion from the JVET test set~\cite{Hanhart2018} at a resolution of $2216 \times 1108$ pixels.
In accordance with the common test conditions for 360-degree video (360-CTC)~\cite{Hanhart2018}, all tests are performed on 32 frames of each sequence using four quantization parameters $\text{QP} \in \{ 22, 27, 32, 37 \}$ with the random access (RA) configuration~\cite{Bossen2020}.
Rate savings are calculated according to the Bjøntegaard Delta (BD) model~\cite{Bjontegaard2001} with respect to PSNR and WS-PSNR~\cite{Sun2017}, which are calculated using the 360Lib software 360Lib-12.0~\cite{Ye2020a, 360Lib-12.0}.

Table~\ref{table:models-table} shows a comparison of the BD-Rates between our proposed MPA and existing 360-degree motion models with respect to VTM-14.2.
Thereby, 3DT denotes the 3D-translational~\cite{Li2017, Li2019}, TAN denotes the tangential~\cite{DeSimone2017}, ROT denotes the rotational~\cite{Vishwanath2017, Vishwanath2018a}, GED denotes the geodesic~\cite{Vishwanath2018}, and TARO denotes the combined tangential and rotational motion model~\cite{Marie2021}.
All motion models are integrated into VVC equivalently to MPA as described in Section~\ref{subsec:codec} including the $4 \times 4$ subblock processing and the compatibility to all listed tools.
The global camera motion vector required by GED is provided as prior information.

With average rate savings of 1.72\% based on PSNR and 1.55\% based on WS-PSNR, MPA shows the best overall performance among the set of 360-degree motion models.
As the closest competitor, GED shows average rate savings of 1.21\% and 1.09\% based on PSNR and WS-PSNR, respectively.
For sequences where the underlying 3D motion is less aligned to the available motion planes, such as \textit{Landing2} and \textit{BranCastle2}, GED is able to outperform MPA.
Nonetheless, MPA shows a highly competitive performance also in these suboptimal scenarios where it still achieves notable rate savings outperforming all other motion models.

\begin{figure}[t!]
  \centering
  \newlength{\framewidth}
  \setlength{\framewidth}{0.292\linewidth}
  \subfloat[Original\label{fig:mpa-preds:1}]{%
  \includegraphics[width=\framewidth]{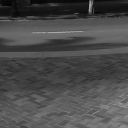}}
  \hfill
  \subfloat[VTM-14.2\label{fig:mpa-preds:2}]{%
  \includegraphics[width=\framewidth]{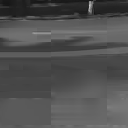}}
  \hfill
  \subfloat[MPA-VVC\label{fig:mpa-preds:3}]{%
  \includegraphics[width=\framewidth]{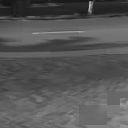}}
  \caption{Crops from original and predicted frame 16 of sequence Broadway at QP32 for VTM-14.2 and the proposed MPA. Best to be viewed enlarged on a monitor.}
  \label{fig:mpa-preds}
\end{figure}

Fig.~\ref{fig:mpa-preds} gives an explanation for the observed rate savings by showing a section of a predicted frame for VTM-14.2 and the proposed MPA-VVC.
For reference, the corresponding section from the original frame is shown as well.
Especially on the paved street surface, it is clearly visible that MPA is able to retain textures at a higher level of detail with less distracting blocking artifacts than VTM-14.2.

\vspace*{-1.2em}\section{Conclusion}\label{sec:conclusion}\vspace*{-0.3em}

In this paper, we proposed a novel motion plane adaptive motion model for spherical video coding that allows to perform inter prediction on different motion planes in 3D space.
Our proposed integration of MPA into the state-of-the-art H.266/VVC video coding standard shows significant BD-rate savings of more than 1.7\% based on PSNR and more than 1.5\% based on WS-PSNR on average.
Our model shows to be highly competitive among related motion models for spherical video coding.
In future work, we plan to extend our approach through a multi-model coding concept, such that the encoder can adaptively select the motion model that leads to the best compression efficiency for each CU.

\bibliographystyle{IEEEbib}
\bibliography{ms}

\end{document}